\documentclass[twocolumn,aps,prd,showpacs,amssymb,floatfix]
{revtex4}
\usepackage{subfigure}
\usepackage[dvips]{graphicx}
\usepackage{mathptmx}      % use Times fonts if available on your TeX system
\usepackage{bm}

\begin{document}

\def\be{\begin{equation}}
\def\ee{\end{equation}}
\def\bea{\begin{eqnarray}}
\def\eea{\end{eqnarray}}
\def\rra{\right\rangle}
\def\lla{\left\langle}
\def\eps{\epsilon}
\def\sgm{\Sigma^-}
\def\la{\Lambda}
\def\pv{\bm{p}}
\def\kv{\bm{k}}
\def\zv{\bm{0}}
\def\bc{B=90\;\rm MeV\!/fm^3}
\def\ms{M_\odot}

\title{Structure of hybrid protoneutron stars within the Nambu--Jona-Lasinio
 model}

\author{G. F. Burgio and S. Plumari}
\affiliation{
INFN Sezione di Catania and 
Dipartimento di Fisica e Astronomia, Universit\`a 
di Catania,\\
Via Santa Sofia 64, 95123 Catania, Italy}

\date{\today}

\begin{abstract}
We investigate the structure of protoneutron stars (PNS) formed by hadronic and
quark matter in $\beta$-equilibrium described by appropriate equations 
of state (EOS). For the hadronic matter, we use a finite temperature EOS 
based on the Brueckner-Bethe-Goldstone many-body
theory, with realistic two- and three-body forces. For the quark sector,
we employ the Nambu--Jona-Lasinio model.
We find that the allowed maximum masses are comprised in a narrow range 
around 1.8 solar masses, with a slight dependence on the temperature.
Metastable hybrid protoneutron stars are not found.
\end{abstract}

\maketitle

%==============================================================================
\section{Introduction}

Protoneutron stars are generally believed to be the evolutionary 
endpoints of the gravitational collapse of a massive star with mass
larger than 8 solar masses \cite{shapiro,bethe}. Several different stages
may happen during the evolution process \cite{burr,praka}. 
Initially, a PNS is very hot, with an entropy per baryon of the order
of 1 to 2, and contains neutrinos produced by electron capture in 
the beta-equilibrated matter. Due to their short mean free paths, they are 
prevented from leaving the star, and are temporarily trapped in it. 
The number of leptons per baryons that remain trapped is approximately 0.4.
The subsequent evolution of the PNS, on a timescale of 10-20 s, is dominated 
by neutrino diffusion and cooling, and the newly formed neutron star (NS) 
stabilizes at practically zero temperature. 

In a previous article \cite{proto} we have studied static properties 
of PNS assuming that nucleons, hyperons, and leptons are present 
in stellar matter. Our calculations are based on the EOS
derived within the Brueckner-Bethe-Goldstone (BBG) theory of nuclear matter
\cite{book}, 
extended to finite temperature. It turned out that for the heaviest PNS, close
to the maximum mass (about two solar masses), the central particle density
reaches values larger than $1/\rm fm^{3}$, thus making possible a hadron-quark 
phase transition. This could have strong consequences for the
dynamical evolution of a PNS into a NS \cite{cooke}. 

Therefore, we have extended the previous calculations, and studied 
hybrid protoneutron stars with a quark core described with the MIT bag model,
finding  maximum masses below 1.6 solar masses, independently on the 
temperature. Moreover, no metastable hybrid PNS were found \cite{hyb06}.

In this paper, we discuss the structure of hybrid PNS using 
the BBG EOS for describing the hadronic phase, 
and the Nambu--Jona-Lasinio model at finite temperature for the quark matter 
(QM) phase. We find that no 
phase transition to quark matter can take place if hyperons are included in the
hadronic phase, with and without neutrino trapping. Thus, the hadronic phase 
contains only nucleons, and as a consequence, neutrino trapping  
will soften the equation of state, and the PNS masses will be smaller
than the NS ones.

We also find that the presence of QM 
limits the value of the maximum mass in a narrow interval around 1.8 $\ms$, 
with a slight dependence on the temperature. 

This paper is organized as follows. In section \ref{s:bhf} we review
the baryonic EOS in the finite-temperature 
Brueckner-Hartree-Fock approach.
Section \ref{s:qm} concerns the QM EOS according to the
Nambu--Jona-Lasinio model, whereas section \ref{s:max} contains results
about the hadron-quark phase transition. 
In section \ref{s:res} we present the results
regarding PNS structure, obtained combining the baryonic and
QM EOS for beta-stable nuclear matter. Finally, in Section \ref{s:end}
we draw our conclusions.\\

%==============================================================================
\section{Hadronic matter equation of state}
\label{s:bhf}

%------------------------------------------------------------------------------
\subsection{The BBG theory at finite temperature}

We start with the description of the hadronic phase. The EOS is based on the
non-relativistic Brueckner-Bethe-Goldstone (BBG) many-body theory \cite{book}, 
which is a linked cluster 
expansion of the energy per nucleon of nuclear matter, well convergent 
\cite{thl} and accurate enough in the density range relevant for neutron stars.

At finite temperature, the formalism which is closest to the BBG expansion, 
and actually reduces to it in the zero-temperature limit, is the one formulated
by Bloch and De Dominicis in \cite{bloch}. This  
has been applied successfully to the study of the limiting temperature
in nuclei \cite{nic}.In this approach the
essential ingredient is the two-body scattering matrix $K$, which,
along with the single-particle potential $U$, satisfies the
self-consistent equations
{\bea
  \langle k_1 k_2 | K(W) | k_3 k_4 \rangle
 &=& \langle k_1 k_2 | V | k_3 k_4 \rangle
\nonumber\\&& \hskip-32mm %27 
 +\; \mathrm{Re}\!\sum_{k_3' k_4'}
 \langle k_1 k_2 | V | k_3' k_4' \rangle
 \frac{ [1\!-n(k_3')] [1\!-n(k_4')]}{W - E_{k_3'} - E_{k_4'} + i\epsilon }
%\nonumber\\&& \hskip-12mm \times
 \langle k_3' k_4' | K(W) | k_3 k_4 \rangle
\nonumber\\&&
\label{eq:kkk}
\eea}
and
\be
 U(k_1) = \sum_{k_2} n(k_2) \langle k_1 k_2 | K(W) | k_1 k_2 \rangle_A \:,
\label{eq:ueq}
\ee
where $k_i$ generally denote momentum, spin, and isospin. 
$W = E_{k_1} + E_{k_2}$ represents the starting energy, 
$E_k = k^2\!/2m + U(k)$ the single-particle energy,
and $n(k)$ is a Fermi distribution. 
For the two-body interaction $V$, we choose the Argonne $V_{18}$
nucleon-nucleon potential \cite{v18}.
We have also introduced three-body forces
(TBF) among nucleons, adopting the phenomenological Urbana model \cite{uix}.
This allows to reproduce correctly the nuclear matter saturation point 
$\rho_0 \approx 0.17~\mathrm{fm}^{-3}$, $E/A \approx -16$ MeV, and gives 
values of incompressibility and symmetry energy at saturation compatible 
with those extracted from phenomenology \cite{myers}. For details, the reader
is referred to Ref.\cite{bbb}. 

The calculation of stable configurations in PNS requires the knowledge
of the EOS at different chemical compositions.
In order to simplify the numerical procedure, we have introduced the so-called
{\em Frozen Correlations Approximation}, 
i.e., the correlations at $T\neq 0$ are assumed to be essentially 
the same as at $T=0$. 
This means that the single-particle potential $U_i(k)$ for the component $i$ 
can be approximated by the one calculated at $T=0$.
It has been shown in ref.~\cite{book} that this assumption is valid with good 
accuracy if the temperature is not too high. 
Within this approximation, for given density and temperature, 
Eqs.~(\ref{eq:kkk}) and (\ref{eq:ueq}) have to be
solved self-consistently along with the equations for the
densities, $\rho_i=\sum_k n_i(k)$,
and the free energy density, 
which has the following simplified expression
\be
 f = \sum_i \left[ \sum_{k} n_i(k)
 \left( \frac{k^2}{2m_i} + \frac{1}{2}U_i(k) \right) - Ts_i \right] \:,
 \label{e:fb}
\ee
where
\be
 s_i = - \sum_{k} \Big( n_i(k) \ln n_i(k) + [1-n_i(k)] \ln [1-n_i(k)] \Big)
\ee
is the entropy density for component $i$ treated as a free gas with
spectrum $E_i(k)$. 
In recent years, the BBG approach at zero temperature has been extended to 
the hyperonic sector
in a fully self-consistent way \cite{hypmat,hypns}, by including the
$\Sigma^-$ and $\Lambda$ hyperons. For that, 
we have used the Nijmegen soft-core nucleon-hyperon (NH) 
potentials NSC89 \cite{maessen}, 
and neglected the hyperon-hyperon (HH) interactions, 
since so far no reliable HH potentials are available.
We have found that 
the presence of hyperons strongly softens the EOS, and produces a
maximum NS mass that lies slightly below the canonical
value of 1.44 $\ms$ \cite{taylor}. 
However, since the quantitative effects of more reliable NH and HH potentials
haven't been explored yet, in this paper we discuss 
finite temperature calculations by using non-interacting hyperons.
This approximation could not be justified for the present
work, because the hadron-quark phase transition occurs at densities much above
the normal nuclear matter saturation density, where hyperons are expected
to play a role. A more complete study with the inclusion of interacting 
hyperons at finite temperature is in progress.

For stars in which the strongly interacting particles are only baryons, 
the composition at given baryon density $\rho$ is determined by imposing
electric charge neutrality and equilibrium under the weak processes
\be
 B_1 \rightarrow B_2 + l + {\overline \nu}_l \ ,\quad   
 B_2 + l \rightarrow B_1 + \nu_l \:,
\label{weak:eps}
\ee
where $B_1$ and $B_2$ are baryons and $l$ is a lepton, 
either an electron or a muon. When the neutrinos are trapped, 
these two requirements imply that the relations
\be
\sum_i q_i x_i + \sum_l q_l x_l = 0
\ee
and
\be
\mu_i = b_i \mu_n - q_i( \mu_l - \mu_{\nu_l}) \:,
\label{e:mutrap}
\ee
are satisfied. In the expression above, $x_i=\rho_i/\rho$ is the baryon 
fraction of the species $i$,
$b_i$ the baryon number, and $q_i$ the electric charge.
Equivalent quantities are defined for the leptons $l=e,\mu$.
The initial PNS contains trapped neutrinos produced in electron capture 
process, so the electron and muon lepton numbers
are conserved on dynamical time scales. We fix the electron lepton number 
$Y_e = x_e+x_{\nu_e} = 0.4$, 
as indicated by gravitational collapse calculations, and
$Y_\mu = x_\mu-x_{\bar \nu_\mu} = 0$,  since no muons are present when 
neutrinos become trapped.

Hence, the composition of beta-stable and charge-neutral baryonic matter is 
determined by the baryon chemical potentials for each species $i$, 
which are related to the free energy density 
\be
 \mu_i = \frac{\partial f}{\partial \rho_i} \:,
\label{e:mufree}
\ee
Therefore, one needs to know the functional dependence of the free energy,
Eq.~(\ref{e:fb}), on the individual partial densities $\rho_i$ and on the 
temperature.
In Ref.~\cite{proto} we have provided analytical parametrizations
of our numerical results for symmetric and neutron matter, from which one can
readily obtain the nucleon chemical potentials in beta-stable matter.
The chemical potentials of the noninteracting hyperons and leptons 
are obtained by solving numerically the free
Fermi gas model at finite temperature.

\begin{figure}[t] %.....................................
\centering 
\includegraphics[width=5.5cm,angle=270]{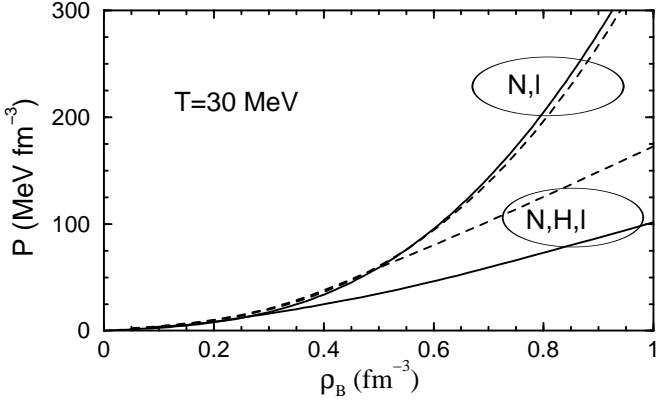}
\caption{
The EOS in beta-stable baryon matter is displayed for temperature 
$\rm T=30~ MeV$. 
The upper curves display results for matter containing nucleons
and leptons, whereas the lower curves are for matter with hyperons.
Solid (dashed) lines refer to the neutrino-free (neutrino-trapped) case.
}
\label{f:eos}
\end{figure}

Once the composition of the $\beta$-stable, charge neutral stellar matter 
is known,
one can calculate the total free energy density $f$ and then the
pressure $p$ through the usual thermodynamical relation
\be
 p = \rho^2 \frac{\partial{(f/\rho)}}{ \partial{\rho}} \:.
\ee
The resulting EOS is displayed in Fig.~\ref{f:eos}, where the
pressure for beta-stable matter, without (solid lines) 
and with (dashed lines) neutrinos, is plotted as a
function of the baryon density at temperature $\rm T=30~MeV$.
The upper curves represent the EOS for stellar matter containing nucleons only,
whereas the lower curves display the case when free hyperons are included.
We notice that if only nucleons and leptons are present in the neutrino-trapped
matter,
the EOS gets softer because nuclear matter is more symmetric.
The presence of hyperons introduces additional softening. However,
the degree of softening is smaller in the neutrino-trapped case
than in the neutrino-free case, because the hyperons
appear at larger densities in neutrino-trapped matter, and 
their concentration is smaller. We have checked that this trend depends
weakly on the temperature, which plays a role only in the low density 
range.

 %.................................................................

%==============================================================================
\section{Quark matter equation of state}
\label{s:qm}

%------------------------------------------------------------------------------
\subsection{The Nambu--Jona-Lasinio model}

We describe the quark phase within the Nambu--Jona-Lasinio model 
\cite{huef,buballa}. The SU(3) version of the model includes most of
symmetries of the QCD Lagrangian, including the spontaneous breakdown of
chiral symmetry, which is essential in treating the lightest hadrons.
The NJL model assumes deconfined point-like quarks, and is not renormalizable,
requiring regularization through a cutoff in three momentum space. 
In its commonly used form, the Lagrangian reads 
\begin{eqnarray}
{\cal{L}} &=& \bar{q}(i\gamma^\mu \partial_{\mu} - \hat m) q + G \sum_{a=0}^8 
\left[ (\bar{q} \tau_a q)^2 + (\bar{q} i \gamma_5 \tau_a q)^2 \right] \nonumber \\  
&- & K \left\lbrace {\rm det}_f [\bar{q} (1+\gamma_5) q] + {\rm det}_f
[\bar{q} (1-\gamma_5) q] \right\rbrace,
\end{eqnarray} 
\noindent
where $q$ are the quark fields with three flavors and three colors, 
and $\tau_a$ are the U(3) flavor matrices. The model parameters are 
the current quark mass matrix $\hat m = {\rm diag}(m_u, m_d, m_s)$,
the coupling constants $G$ and $K$, and the cutoff in the three-momentum
space, $\Lambda$.
Following ref.\cite{huef}, we choose $\Lambda=602.3~ \rm MeV$, the two 
coupling constants $G=1.835/\Lambda^2$ and $K=12.36/\Lambda^5$, as well as 
the current quark masses $m_u=m_d=5.5 ~\rm MeV$, $m_s=140.7 ~\rm MeV$.
The values chosen for those 
parameters have been adjusted to reproduce masses and decay constants of the 
pseudoscalar meson nonet.

In the mean-field approximation at finite temperature and chemical potential, 
the thermodynamical potential of the quark phase is
$\Omega = \Omega_{\rm FG} + \Omega_{\rm Int}$, where

\be
\frac{\Omega_{\rm FG}}{V}(T,\mu_i) = 2 N_c T
\sum_{i=u,d,s} \int \frac{d^3p}{(2\pi)^3} \left[{\rm ln}(1-f_i)+ {\rm ln}
(1- \tilde f_i) \right]
\ee
\noindent 
is the Fermi gas contribution arising from quarks. 
We consider three flavors and three colors, hence $N_c = 3$.
The distribution functions of fermions and anti-fermions are given by
$f_i = [1+{\rm exp}(\frac{1}{T} (E_i - \mu_i))]^{-1}$ and
$\tilde f_i= [1+{\rm exp}(\frac{1}{T} (E_i + \mu_i))]^{-1}$.
$E_i$ and $\mu_i$ denote the single particle energy and chemical potential of 
the quark flavor $i$. 

The thermodynamical potential due to interactions among quarks is given by
%\begin{widetext}
\begin{eqnarray}
\frac{\Omega_{\rm Int}}{V}(T, \mu_i) &=& -2 N_c \sum_{i=u,d,s} \int \frac{d^3p}{(2\pi)^3} 
\left( \sqrt{M_i^2 + p^2} - \sqrt{m_i^2 + p^2} \right ) \nonumber \\
& + & 2G  <\bar q_i q_i>^2 \nonumber \\
& - & 4K <\bar q_u q_u> <\bar q_d q_d>  <\bar q_s q_s>
\label{e:om_int}
\end{eqnarray}
%\end{widetext}
\noindent
In the NJL model, the quark masses are dynamically generated as solutions of
the gap equation,
obtained by imposing that the potential be stationary with respect to 
variations in the quark condensate $<\bar q_i q_i>$,
thus finding

\be
M_i = m_i -4G  <\bar q_i q_i>^2  + 2K <\bar q_j q_j>  <\bar q_k q_k>
\label{e:gap}
\ee
being $(q_i,q_j,q_k)=\rm {any~permutation~of~} \it {(u,d,s)}$. 
The quark condensate $<\bar q_iq_i>$ and the quark number density $n_i$ are 
given respectively by

\bea
<\bar q_iq_i> &=& -2 N_c \int  \frac{d^3p}{(2\pi)^3} \frac{M_i}{E_i}
\left [ 1 - f_i - \tilde f_i \right ], \nonumber \\
n_i &=&  2 N_c \int  \frac{d^3p}{(2\pi)^3} 
\left [ f_i - \tilde f_i \right ]. 
\label{e:cond}
\eea
\noindent
Eq.(\ref{e:cond}) has to be evaluated self-consistently with Eq.(\ref{e:gap}),
forming a set of six coupled equations for the constituent masses $M_i$. 
Once the self-consistent solutions are
found, we can calculate the pressure and the energy density through the usual 
thermodynamical relations 
\be
p=-\Omega, \quad\quad \epsilon=\Omega + T s + \sum_i \mu_i n_i
\ee
where $s=\partial \Omega / \partial T$ is the entropy density and
$n_i = -\partial \Omega/\partial \mu_i$ are the number densities of flavor $i$.
The total quark number density and the baryon number density are given by
$n=\sum_i n_i$ and $\rho_B=n/3$, respectively. The reader can find more details
in ref.\cite{buballa}.

In a PNS with quark matter and trapped neutrinos we must 
impose beta equilibrium, charge neutrality, and baryon and lepton number 
conservation.
More precisely, the individual quark chemical potentials
are fixed by Eq.~(\ref{e:mutrap}) with $b_q=1/3$, 
which implies
\be
 \mu_d = \mu_s = \mu_u + \mu_l - \mu_{\nu_l} \:.
\label{eq:chem} 
\ee
The charge neutrality condition and the total
baryon number conservation read
\bea
n_e + n_\mu &=& \frac{1}{3}(2 n_u - n_d - n_s) \:,
\\
 3 \rho_B &=& n_u + n_d + n_s.
\label{eq:baryon} 
\eea
These equations determine the composition
and the pressure of the quark matter phase. 

\begin{figure}[h] %......................................
\centering
\includegraphics[width=10cm,angle=270]{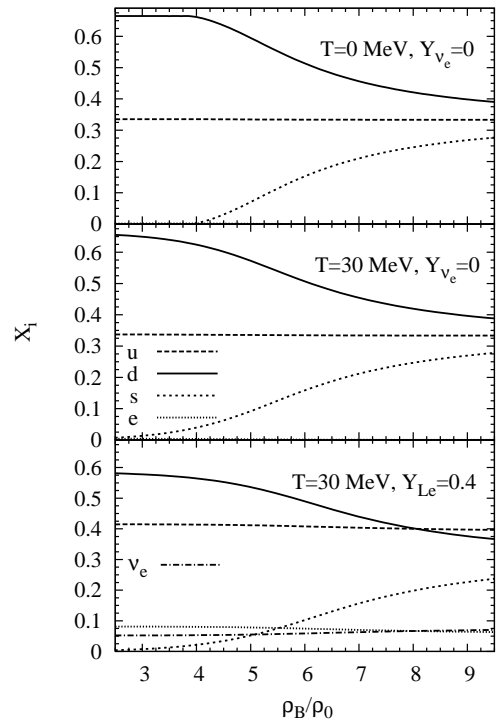}
\caption{ 
The fraction of $u,d,s$ quarks is displayed 
as a function of baryon density for several cases, i.e. $T=0$ (upper panel), and 
$T=30$ MeV  (middle panel) without neutrino trapping. The case for $T=30$ MeV with 
neutrino trapping is shown in the lower panel.} 
\label{f:pop_q}
\end{figure}

\begin{figure}[ht] %........................................................
\centering
\includegraphics[width=6.5cm,angle=270]{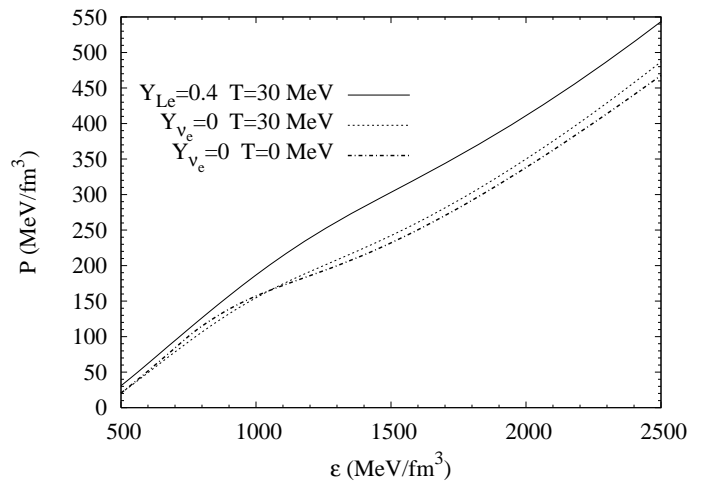}
\caption{ 
Pressure as a function of the mass-energy density for beta-stable, charge
neutral quark matter. See text for details.} 
\label{f:pc}
\end{figure}

Let us discuss first the main features of beta-equilibrated and charge neutral 
quark matter within the NJL model. 
In Fig.~\ref{f:pop_q} we show particle fractions $x_i=\rho_i/\rho$ as a 
function of the baryon density for
neutrino-free (upper and middle panels), and neutrino-trapped quark matter 
(lower panel). At zero temperature, we observe a substantial amount 
of $u$ and $d$ quarks, whereas the $s$-quark starts to appear at baryon 
density four times larger than the normal saturation value. With increasing 
density, the $s$-quark population keeps smaller than the $u$ and $d$ 
populations. With increasing temperature, we observe a slight change of the 
particle population mainly at low density, because of the tails in the Fermi 
distributions. In this case, the onset for $s$-quarks takes place 
at density smaller than in the cold case. The presence of neutrinos 
influences quite strongly the composition, as displayed in 
the lower panel of Fig.~\ref{f:pop_q}. In this case the relative
fraction of $u$ quarks increases substantially from $33\%$ to about
$42\%$, compensating the charge of the electrons that are present at
an average percentage of $8\%$ throughout the considered 
range of baryon density, whereas the $d$ and $s$ quark fractions decrease.
This implies that in the neutrino trapped case, being
the  strangeness content smaller than in the neutrino-free case, the equation
of state for quark matter will be stiffer. 
This is indeed shown in Fig.~\ref{f:pc}, where the pressure
is displayed as a function of the mass-energy density for the neutrino-free
(lower curves), and for neutrino-trapped case (upper curve). The behaviour
of the pressure is consequence of the quark population in beta-stable matter. 
We clearly observe a kink in the pressure, 
corresponding to the onset of the $s$-quark. The kink becomes smoother in 
the neutrino-trapped case, because the $s$-quark concentration is smaller.
Moreover, the 
EOS for $\beta$-stable, charge neutral quark matter derived within the 
NJL model shows a clear dependence on the trapped lepton fraction. 
This is different from the EOS derived with the MIT bag model for describing 
the quark phase, where no dependence on the trapped lepton fraction is
actually found \cite{hyb06,mene}.

 %........................................................

 %................................................................

%------------------------------------------------------------------------------
\section{Phase transition in hot beta-stable matter}
\label{s:max}

In order to perform the hadron-quark phase transition in 
beta-stable matter at finite temperature, we
adopt the simple Maxwell construction. The more general Gibbs
construction \cite{gle,glen} is still affected by many theoretical 
uncertainties \cite{mixed}, and in any case the final mass-radius 
relation of massive (proto)neutron stars \cite{mit} is slightly affected.

\begin{figure}[t] %............................................................
\centering
\includegraphics[width=10cm,angle=270]{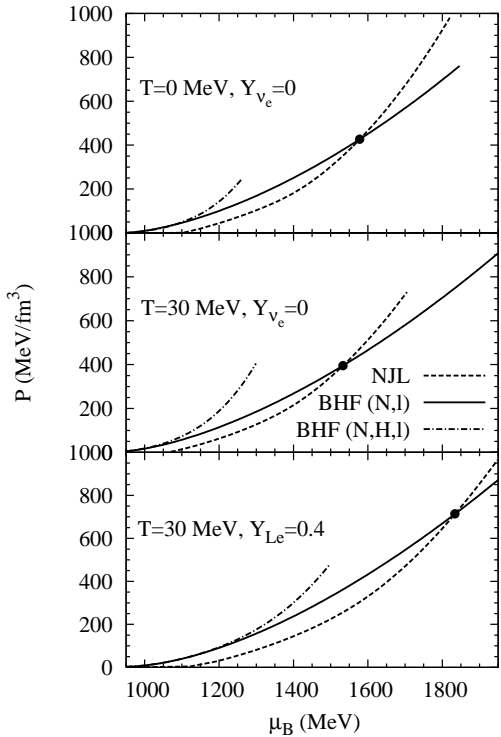}
\caption{ 
Pressure as a function of the baryon chemical potential for beta-stable, charge
neutral nuclear matter and quark matter. See text for details.} 
\label{f:pmu}
\end{figure}

Assuming a first-order phase transition, we impose thermal, chemical, and
mechanical equilibrium between the two phases. This implies
that the phase coexistence is determined by a crossing point in the 
pressure vs. chemical potential plot, as shown in Fig.~\ref{f:pmu}.
There we display the pressure
$p$ as function of the baryon chemical potential $\mu_B$ for 
baryonic and quark matter phases at temperatures $T=0,~30$ MeV.
In the upper panel we show the case at $T=0$ without neutrinos. 
The solid line represents the calculations performed with the BBG method with
only nucleons, the dot-dashed line displays the case when free hyperons are 
included, and the dashed line is the NJL calculation. We observe that the 
phase transition, marked by a full dot, occurs if only nucleons are present
in baryonic matter, as it was already found in ref.\cite{njl}. 
In the BHF approach, this is due to the strong softening of the baryonic 
equation of state when hyperons are included. The same holds true at finite 
temperature, without and with neutrino trapping, as clearly shown in 
Fig.~\ref{f:pmu}. 
In the middle panel, we display results at $T=30$ MeV 
without neutrino trapping.
We observe that the crossing point is only slightly affected by thermal 
effects, and is shifted towards smaller values of the chemical potential. 
This means that the phase transition starts at a baryon density smaller 
than in the cold case. Finally, in the lower panel, we display the 
neutrino-trapped case at $T=30$ MeV.
Neutrino trapping makes the baryonic EOS softer and the quark matter EOS 
stiffer, hence the phase transition is shifted at larger values of the
chemical potential, and of the baryon density.

 %................................................................

\begin{figure}[t] %............................................................
\centering
\includegraphics[width=6.5cm,angle=270]{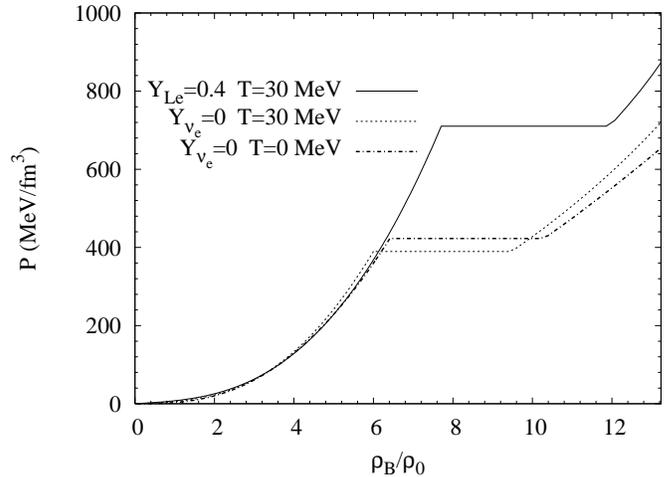}
\caption{ 
Pressure as a function of the baryon density, normalized with respect to
the nuclear matter saturation density $\rho_0$.} 
\label{f:prho}
\end{figure}

In Fig.~\ref{f:prho} we display the pressure as a function of the 
baryon density for the several cases discussed above. The plateaus are
consequence of the Maxwell construction. Below the plateau, 
$\beta$-stable and charge neutral stellar matter is in the purely 
hadronic phase, whereas for density above the ones characterizing the 
plateau, the system is in the pure quark phase.
We notice the neutrino trapping shifts the quark matter onset at larger
values, and gives rise to a wider plateau. 

Similar calculations have been performed in ref.\cite{provi}, 
where the non linear Walecka model \cite{sw} has been adopted for the hadronic 
phase, which includes nucleons and hyperons, and the NJL model with a 
different choice of the parameters for the description of the quark phase. 
The phase transition between the two phases has been performed 
adopting the Gibbs construction. Also in this case, the onset of quark matter 
appears at values of the baryon density which decrease with increasing the 
temperature. If neutrino trapping is taken into account, the onset of quark 
matter is shifted at larger values of the baryon density. 
In general, the larger the temperature, the smaller is the density of quark 
matter onset. We notice that, in spite of the different theoretical framework, 
we obtain quite similar results.

 %................................................................

\begin{figure*}[t] %........................................................
\centering
\includegraphics[width=7cm,angle=270]{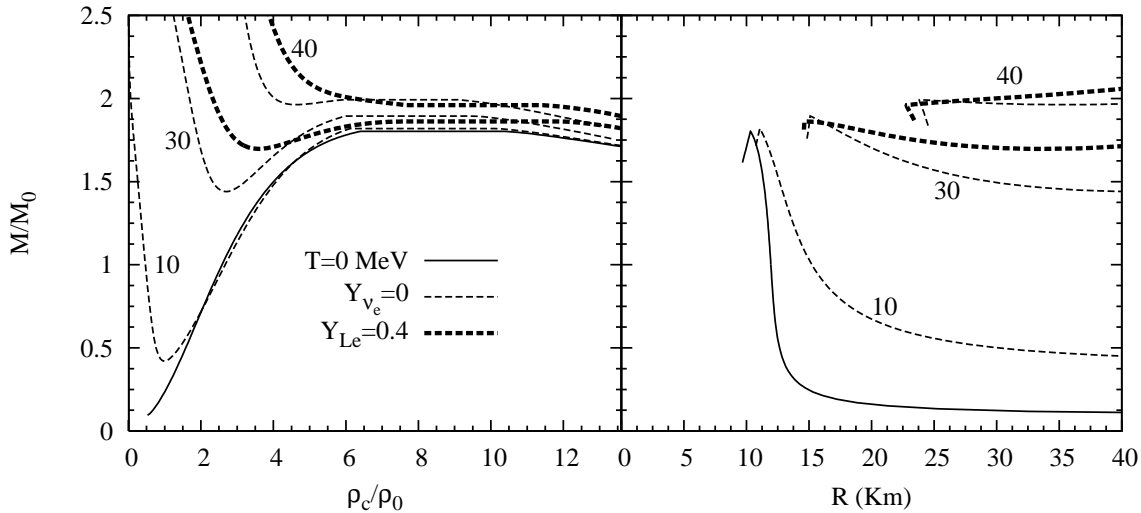}
\caption{ 
The mass-central density (radius) relation is displayed in the left (right) 
panel. The mass is given in units of the solar mass $M_\odot = 
\rm 1.989 \times 10^{33} g$, and the central density $\rho_c$ is normalized 
with respect to the saturation value $\rho_0=\rm 0.17~fm^{-3}$.} 
\label{f:mc}
\end{figure*} %................................................................

%==============================================================================
\section{Protoneutron star structure}
\label{s:res}
Based on these results for the beta-stable baryon and QM phases, 
we can now determine the properties of static (proto)-neutron stars.
The stable configurations can be obtained
from the well-known Tolman-Oppenheimer-Volkoff (TOV) equations \cite{shapiro}
for the pressure $p$ and the enclosed mass $m$, 
\bea
  \frac{dp(r)}{ dr} &=& -\frac{ G m(r) \epsilon(r)}{ r^2 }
\nonumber
  { \left[ 1 + \frac{p(r)}{\epsilon(r)} \right]
  \left[ 1 + \frac{4\pi r^3 p(r)}{m(r)} \right]
  \over
  1 - {2G m(r)/ r} } \:, \\ 
 {dm(r) \over dr} &=& 4 \pi r^2 \epsilon(r) \:,
\label{e:tov}
\eea
once the EOS $p(\epsilon)$ is specified, being $\epsilon$ the total energy 
density ($G$ is the the gravitational constant). Starting with a central
energy density $\epsilon(r=0) \equiv \epsilon_c$, the numerical integration
of eqs.(\ref{e:tov}) provides the mass-radius relation.

Simulations of supernovae explosions \cite{burr,pons} show that
during the early stage, the entropy
profile decreases from the surface to the core starting from values
of 6--10 \cite{burr}, and the temperature
drops rapidly to zero at the surface of the star due to the fast
cooling of the outer part of the PNS, where the stellar matter is
transparent to neutrinos. Therefore, we have modelled the PNS by assuming
a hot isothermal and neutrino-opaque core separated from an outer cold crust
\cite{bps,fmt}
by an isentropic, beta-equilibrated, and neutrino-free intermediate layer
described by the Lattimer and Swesty equation of state \cite{lat}.
For details, the reader is referred to ref.\cite{hyb06,nicot1}.

In the following we consider two snapshots which represent the 
thermodynamical conditions in an evolving PNS : 
i) the initial state consisting of a PNS with a hot 
($T_{\rm core}\approx$ 30--40 MeV) 
neutrino-trapped core and a high-entropy transition layer
($S_{\rm env.}\approx$ 6--8),
joined to a cold outer crust; ii) the final state which 
represents the short-term cooling, 
where the neutrino-free core possesses a low temperature of about $10$ MeV 
and is direcly attached to a cold crust.

The results are plotted in Fig.~\ref{f:mc}, where we
display the gravitational mass $M_G$ (in units of the solar mass
$\ms$) as a function of the central
baryon density $\rho_c$ (left panels) and the radius $R$ (right panels).
We display the complete set of results at core
temperatures $T=0,~10~ \rm MeV$ without neutrino
trapping, and $T=30,~40~\rm MeV$ without and with neutrino trapping.
Due to the Maxwell construction, the curves are not continuous and 
display plateaus in the 
mass-central density relation, which disappear in the case of
the Gibbs construction. For values of central density smaller than the one 
characterizing the maximum mass, the PNS are purely baryonic stars. 
Then an increase of central density is required in order to start the quark 
phase, as shown by the phase diagram of Fig.\ref{f:prho}. 
We find that the onset of the pure quark phase at the center of the PNS as the 
mass increases marks an instability of the star, i.e. the PNS collapses to a 
black hole at the transition point since the quark EoS is unable to sustain the
increasing central pressure due to gravity. We had already found 
the same  results in the case at T=0 \cite{njl}. 
It has been argued that this instability might be related to the lack of 
confinement in the original NJL model \cite{inst}.
Thus, heavy hybrid PNS in this model are practically baryonic stars up to 
large values of the central densities, eventually with a core in a mixed 
hadron-quark phase.
The characteristics of the maximum mass configurations are reported 
separately in Table \ref{t:mass}. 
We notice that the maximum PNS masses are comprised in a small range around
1.8 solar masses, with a slight dependence on the temperature.
Moreover, they turn out to be smaller than those of cold NS.
This is due to the fact that the baryonic phase contains only nucleons,
in which case neutrino trapping softens the EOS. This leads to smaller
maximum masses, and excludes the possibility of metastable configurations.

Below the maximum mass configuration, however, the stars develop
an extended outer envelope of hot matter, 
the details of which depend on the treatment of the low-density 
baryonic phase. The temperature dependence of the curves is quite
pronounced for intermediate and low-mass stars, showing a strong
increase of the minimum mass with temperature.
Above core temperatures of about 40 MeV all stellar configurations become 
unstable.

\begin{table} %......................................................
\begin{center}
\bigskip
\begin{ruledtabular}
\begin{tabular}{l|c|cccc}
 Composition           & $T$ (MeV) & $M_G/M_\odot$ & $R$ (km) & 
$\rho_c/\rho_0$ & $\epsilon_c (\rm MeV/fm^3)$ \\
\hline
                       & 0         & 1.8         & 10.3  & 6.4  & 1294\\
 $N,QM,l$              & 10        & 1.82        & 11.  & 6.12  & 1217  \\
                       & 30        & 1.9        & 15.  &  6.    & 1168 \\
                       & 40        & 1.99        & 23.8  & 5.94  & 1138  \\
\hline
                       & 0         &  1.76        &  9.4   &  7.82 & 1751  \\
 $N,QM, l,\nu_e$       & 10        & 1.78        & 10.9  & 7.76  & 1730\\
                       & 30        & 1.86        & 14.6  & 7.7  & 1689   \\
                       & 40        &  --         &  --   &  --  & --   \\
\end{tabular}
\end{ruledtabular}
\end{center}
\caption{
Properties of the maximum mass configuration for different stellar 
compositions and temperatures.}\label{t:mass}
\end{table} %..................................................................

%==============================================================================
\section{Conclusions}
\label{s:end}

In this article we have studied hybrid PNS, combining the most recent 
microscopic baryonic EOS in the BBG approach at finite temperature 
with the Nambu--Jona-Lasinio model for describing the QM phase.
Within this approach, we have found that the hadron-quark phase transition 
can take place if baryonic matter contains only nucleons. The inclusion of 
free hyperons softens considerably the EOS, and therefore inhibits any 
phase transition to quark matter. This is in agreement with the findings
obtained at $T=0$ in Ref.\cite{njl}, where the Njimegen soft-core 
nucleon-hyperon
interaction was included. This requires new experimental data on hypernuclei, 
from which more modern parametrizations of the nucleon-hyperon and/or 
hyperon-hyperon interaction may be extracted.

Within the NJL model for quark matter, we have also found the following :
i) the transition density to the quark phase in neutrino-free baryonic matter
occurs at about 6 times normal nuclear matter saturation density;  
ii) with increasing temperature, the onset is shifted to 
smaller and smaller values of the baryon density, and
iii) neutrino trapping substantially increases the transition density to QM.

Therefore the hadron-quark phase transition studied with the original NJL 
model possesses general features, which are quite similar to the ones
found with the MIT bag model \cite{hyb06}, 
where an increase of the temperature shifts the transition density to lower 
values. Moreover, in this approach, hyperons are present in 
the hadron-quark mixed phase.

We have also found that maximum mass for hybrid PNS lie in a narrow range
around 1.8 solar masses, and that they are smaller than the 
corresponding ones obtained in the cold, neutrino-free case. Therefore,
metastable configurations are not possible.  This same result was found in
ref.\cite{hyb06}, where the MIT bag model was employed for describing the
quark phase. Both results
are is in contrast to Ref.~\cite{cooke}, where such
metastability was found for hybrid stars, but with a much stiffer baryonic 
EOS including hyperons.

Again, we confirm our prediction of limiting masses for PNS smaller than 2 
solar masses, which are compatible with currently established
observational data on NS. 

%------------------------------------------------------------------------------
\begin{acknowledgments}
We warmly thank  J. Pons (University of Alicante, 
Spain) and M. Baldo (INFN Sezione di Catania, Italy) for fruitful discussions.
\end{acknowledgments}
%------------------------------------------------------------------------------

%%%%%%%%%%%%%%%%%%%%%%%%%%%%%%%%%%%%%%%%%%%%%%%%%%%%%%%%%%%%%%%%%%%%%%%%%%%%%%%

\end{document}